\documentstyle[preprint,prd,aps,amsfonts]{revtex}

\begin{document}
\preprint{DTP 96--35}
\draft
\title{Static solitons with non-zero Hopf number}

\author{Jens Gladikowski${}^a$\thanks{e-mail: 
{\tt Jens.Gladikowski@durham.ac.uk}}
and Meik Hellmund${}^b$\thanks{e-mail: 
{\tt hellmund@tph100.physik,uni-leipzig.de}}}

\address{ ${}^{a)}$Department of Mathematical Sciences,
South Road, Durham DH1 3LE, England\\
${}^{b)}$Institut f\"ur Theoretische Physik, Augustusplatz,
D-04109 Leipzig,Germany}
\date{\today}
\maketitle

\begin{abstract}
We investigate a generalized
non-linear $O(3)$ $\sigma$-model in three
space dimensions where the fields are maps from ${\Bbb R}^3 \cup \{
\infty \}$ to $S^2$.
Such maps are classified by a homotopy invariant 
called the Hopf number which takes integer values.
The model exhibits
soliton solutions of closed vortex type which have a lower
topological bound on their energies.
We numerically compute the fields for topological charge 1 and 2
and discuss their shapes and binding energies.
The effect of an additional potential term is considered and
an approximation is given
for the spectrum of slowly rotating solitons.
\end{abstract}

\pacs{11.27.+d, 11.10.Lm\vspace{2cm}\\
To appear in Phys. Rev. D15, {\bf 56}}

\section{Introduction}

The non-linear $O(3)$ $\sigma$-model in (3+1) dimensional
space-time is a scalar field theory whose target space is
$S^2$. 
The static fields are maps ${\Bbb R}^3 \cup \{\infty \}  \mapsto S^2$
and can be classified by a homotopy invariant which is called
the Hopf number.
Such a model in three space dimensions
must include higher order terms in the field
gradient in order to allow non-singular, topologically non-trivial,
static solutions.
The corresponding ``sigma model with a Skyrme term'' was proposed
long ago by L.D. Faddeev \cite{Fad:76}.
For this model the Hopf number provides a lower topological bound
on the energy \cite{WaK}.

Early studies on ``Hop{f}ions'' (soliton solutions of Hopf number unity)
in classical field theory, including estimates for their size and mass,
were carried out by de~Vega\cite{deV}.
Subsequently it was suggested to employ them in 
an effective chiral theory describing  low-energy hadron dynamics;
in that respect they are similar to Skyrmions \cite{Gip/Tze:80}. 
It was later shown by Kundu and Rybakov \cite{akr}
that Hop{f}ions in the $O(3)$ $\sigma$-model are of closed vortex type.

Models with non-zero Hopf number 
have also been investigated in condensed matter physics
for the description of three-dimensional ferromagnets and
superfluid ${}^3$He\cite{DzI,VoM}.
These are effective theories of Ginzburg-Landau type 
where the fields are interpreted as  physical order parameters.
However, a field configuration which is a solution of the
full equations of motion has not been found for any of the
mentioned theories.

In this paper we  mainly study  classical static Hop{f}ions.
Our model is defined in section \ref{Hmaps} where also
an ansatz of azimuthal symmetry  is introduced which is later 
used for numerical computations.
In section \ref{Nres} we present our numerical results which are
minima of the energy functional for 
Hopf number one and two.
We discuss their shapes and binding energies as well
as their relation to (2+1) dimensional solitons.
Our model has  a  self-interaction
coupling parameter and we study the 
dependence of the energy on this coupling.
In addition, the effect of a symmetry breaking
potential term is described.
In section \ref{Srot} we give a simple
approximation for the excitation spectrum
of a Hop{f}ion slowly  rotating around its axis of symmetry.
We conclude with section \ref{Conc} where we \mbox{also remark on
possible further investigations.}

\section{Hopf maps and toroidal ansatz}\label{Hmaps}

We are almost exclusively interested in static solutions
and therefore define
our model by the following energy functional  on ${\Bbb R}^3$
\begin{equation}
\label{en}
E_{stat}\left[ \bbox{\phi} \right] =
\Lambda\int_{{\Bbb R}^3} d{\bf x}\;
\frac{1}{2}
\left(\partial_i \bbox{\phi} \right)^2 +
\frac{g_1}{8}\left(\partial_i \bbox{\phi}  \times
\partial_j \bbox{\phi} \right)^2 +
\frac{g_2}{8}\left(\partial_i \bbox{\phi} \right)^2
\left( \partial_j \bbox{\phi}\right)^2 \,.
\end{equation}
For $g_2=0$ this is equivalent to the static
energy of the Faddeev-Skyrme model \cite{Fad:76,WaK}.
The field $\bbox{\phi}$ is a three component vector in
iso-space, subject to the constraint $\bbox{\phi}^2=1$.
The cross product is taken in internal space and the coordinate
indices $i,j$ run from 1 to 3.

For $g_1=g_2=0$
minima of $E$ (eq.~(\ref{en})) are harmonic maps from ${\Bbb R}^3$
to $S^2.$ As shown in ref.~\cite{Bai/Woo:88}, all  non-constant
harmonic maps are orthogonal projections 
${\Bbb R}^3 \mapsto {\Bbb R}^2$, followed by a
harmonic map ${\Bbb R}^2 \mapsto S^2$ and therefore have infinite energy.

Consistently,
simple scaling arguments along the line of the Hobart-Derrick theorem
\cite{Hob:63/der} show that the fourth order terms in the energy functional
are required to stabilize the soliton against shrinkage. We  include here
the most general combination of global $O(3)$-invariant fourth order terms.

The parameter $\Lambda$ is a constant of dimension energy/length and 
determines  the models energy unit.
The couplings $g_1$ and $g_2$ are of dimension (length)${}^2.$
The ratio $g_1/g_2$ is the only physically relevant coupling
since an overall scaling of  $g_1$ and $g_2$
can be absorbed by a rescaling of length and energy units.
Using  $\left(\partial_i \bbox{\phi} \times
\partial_j \bbox{\phi} \right)^2$ = $
\left(\partial_i \bbox{\phi}\right)^2\left(\partial_j \bbox{\phi}\right)^2-
\left(\partial_i \bbox{\phi} \cdot \partial_j \bbox{\phi}\right)^2$
and the inequality
\begin{equation}
2\sum_{i j}
\left(\partial_i \bbox{\phi} \cdot \partial_j \bbox{\phi}\right)^2
\geq
\sum_{i j}
\left(\partial_i \bbox{\phi}\right)^2\left(\partial_j \bbox{\phi}\right)^2 
\geq
\sum_{i j}
\left(\partial_i \bbox{\phi} \cdot \partial_j \bbox{\phi}\right)^2\,,
\end{equation}
one sees that the allowed ranges for the coupling constants are
$g_2 \geq 0$ and $g_1 > -2g_2.$
 
For finite energy solutions one requires $\bbox{\phi} \to {\bf n}$
as $\left|{\bf r}\right| \to \infty$, where ${\bf n}$ is a constant
unit vector.
Thus ${\Bbb R}^3$ can be one-point compactified to
$S^3$ and the fields $\bbox{\phi}$ are  maps
\begin{equation}
\bbox{\phi} : \quad S^3 \mapsto S^2.
\end{equation}
Because $\pi_3(S^2)={\Bbb Z}$, every $\bbox{\phi}$ falls into a 
class of topologically equivalent maps, where each class
is characterized by an integer: the Hopf number $H$. 

Although it is not a simple ``winding number'', $H$ has an
elementary  geometric interpretation.
The pre-image of
every point of the target space $S^2$ is isomorphic to a circle.
All  those circles are interlinked with each other in the 
sense that any  circle intersects the disc spanned by  
any other one.
The Hopf number just equals the multiplicity by  which two
arbitrary circles are linked.

$H$ also has a differential geometric representation
~\cite{Bott82}: 
If $f$ is a generator of  the de-Rham cohomology
$H^2_{DR}(S^2)$, its pullback $F$ under
$\bbox{\phi}$ is necessarily exact  since $H^2_{DR}(S^3)=0.$  
Hence a 1-form $A$ with $F = dA$ exist and
$H \sim \int A\wedge F.$

In coordinate language, the dual of $F$ is
$B_i=\varepsilon_{ijk}\,\bbox{\phi}\cdot\partial_j\bbox{\phi} \times
\partial_k\bbox{\phi}$
and
\begin{equation}\label{hopfnr}
H = -\frac{1}{(8\pi)^2}\int_{{\Bbb R}^3} d{\bf x}\;
{\bf B \cdot A}\,.
\end{equation}

It was proved in \cite{WaK} that  
the energy eq.~(\ref{en}) has a lower topological bound 
in terms  of $H$. For $g_1 \geq 0$  it is given by
\begin{equation}\label{topbound}
E_{stat} \geq \Lambda k H^{3/4}\,,
\end{equation}
where $k=\sqrt{2g_1}(2\pi)^23^{3/8}$ \cite{akr}.

The variational equations resulting from eq.~(\ref{en}) are coupled
non-linear partial differential equations.
It would be useful to find a parametrization of $\bbox{\phi}$
which carries  non-zero Hopf charge and
allows the equations to be reduced  to ordinary
differential equations.
There have been two proposals for such fields in the
literature.
One of them uses 
spherical coordinates and is a composition of the standard Hopf map
and a map $S^2 \mapsto S^2$ for which a hedgehog ansatz is employed
\cite{VoM,fot/zhs}.
Alternatively, a closed vortex ansatz in toroidal
coordinates was suggested 
\cite{deV,Gip/Tze:80,Meis:85,Wu/Zee:89}.
However, as shown in \cite{Kund:86},
even for $g_2=0$  none of these proposals
allows a consistent separation of variables in
the variational equations derived from eq.~(\ref{en}).

At this point it is instructive to look at the symmetries of the field.
It was shown in ref.~\cite{akr}
that the maximal subgroup of
$O(3)_X \otimes O(3)_I$ under which 
fields with non-vanishing
Hopf number can be invariant is
\begin{equation}\label{symm}
G = 
\text{diag} \left[ O(2)_X \otimes O(2)_I \right].
\end{equation}
Here $O(2)_X$ and $O(2)_I$ denote rotations
about a fixed axis in space and iso-space respectively.
We choose the $z$-  and $\phi_3$-axis as the axes of symmetry.
According to the Coleman-Palais theorem we expect to find
the minimal energy solution  in the class of
$G$-invariant configurations~\cite{Mak/Ryb:SkMod}. 
Therefore we
use the  most general $G$-invariant ansatz,
written in terms of two functions $w(\xi_1, \xi_2)$ and 
$v(\xi_1,\xi_2).$ They depend on coordinates $\xi_1$ and
$\xi_2$ which form an orthogonal coordinate system
together with  $\alpha$, the angle around the $z$-axis:
\begin{equation}\label{ansatz1}
\phi_1 + i \phi_2 = \sqrt{1- w^2(\xi_1, \xi_2)}
e^{i(N\alpha + v(\xi_1, \xi_2))}
\,, \qquad \phi_3 = w(\xi_1, \xi_2).
\end{equation}
We have checked the consistency of this ansatz with the
variational equations derived from eq.~(\ref{en}).
The components $\phi_1$ and $\phi_2$ have to vanish  along the $z$-axis
for the field to be well-defined.
This is realized
by setting $\bbox{\phi}(0,0,z)$ =
{\bf n} = (0, 0, 1), which 
also defines the vacuum state of the theory.
In order to describe 
a non-trivial map, $\bbox{\phi}$ has to
be surjective.
Hence there is at least one point ${\bf r}_0$ with
$\bbox{\phi}({\bf r}_0) = - {\bf n}$.
Under the action of $G$, ${\bf r}_0$ represents a full circle
around the $z$-axis.
We fix our coordinate system such that this circle lies in the
$xy$-plane and define $a \equiv \left| {\bf r}_0 \right|$.
On every trajectory from the circle to the $z$-axis or
infinity, $w(\xi_1, \xi_2)$ runs at least once from $-1$ to $1.$
Therefore the surfaces of constant $w$  are homeomorphic to tori.

This structure prompts us to choose toroidal  coordinates
$(\eta, \beta, \alpha)$,
related to cylindrical
coordinates $(r,z,\alpha)$ as follows

\begin{equation}
r = \frac{a \sinh \eta}{\tau}, \qquad
z = \frac{a \sin \beta}{\tau}\,,
\end{equation}
where $\tau = \cosh \eta - \cos \beta$.
Surfaces of constant $\eta$ describe  tori about the
$z$-axis, while
each of these tori is parametrized by the  two angles
$(\beta,\alpha)$.
The two cases $\eta =0$ and $\eta =\infty$ correspond to
degenerated tori,
 $\eta = 0$ being the $z$-axis and 
 $\eta = \infty$ the
circle of radius $a$ in the $xy$-plane.

The function $w(\eta,\beta)$ is subject to the boundary conditions
$w(0, \beta) =1, w(\infty, \beta)=-1$ and is periodic in
$\beta.$ 
$v(\eta,\beta)$ is an angle around $\phi_3$ and can include windings
around $\beta.$ Therefore we set $v(\eta,\beta)=M\beta+v_0(\eta,\beta)$ 
where $v_0(.,\beta): S^1\mapsto S^1$ is homotopic to the constant map.
Since $v$ is ill-defined for $w=\pm 1,$ it is not restricted by
any boundary condition at $\eta=0,\infty.$

The ``potential'' ${\bf A}$ and the ``field strength''
${\bf B}$ for this ansatz  are given by

\begin{eqnarray}\label{AB}
A_\alpha = 2\frac{\tau}{a \sinh\eta}N(w-1),
\qquad
A_\beta = 2\frac{\tau}{a}(M+\dot{v}_0)(w+1),\qquad
A_\eta = 2\frac{\tau}{a} v_0^\prime(w+1),\nonumber\\
B_\alpha = 2\frac{\tau^2}{a^2}
( w^\prime(M+ \dot{v}_0) -  v_0^\prime\dot{w}),\qquad 
B_\beta = -2\frac{\tau^2}{a^2\sinh\eta}N w^\prime,\qquad
B_\eta = 2\frac{\tau^2}{a^2\sinh\eta}N
\dot{w},
\end{eqnarray}
where  the dot and prime denote derivatives
with respect to $\beta$ and $\eta$ respectively.
Note that the field $\bf A$ is  well defined on all of ${\Bbb R}^3.$
 The gauge has been chosen such that 
$A_\alpha$ vanishes for $\eta=0$ (where the coordinate $\alpha$ 
is undefined) and
analogously $A_\beta$ vanishes for $\eta=\infty.$

Eq.~(\ref{hopfnr}) then gives
$H=N\,M$ in accordance with the linking number
definition given above.
The energy eq.~(\ref{en}) of ansatz eq.~(\ref{ansatz1}) is given by

\begin{eqnarray}\label{en2}
E[w(\eta,\beta),v(\eta,\beta),a] & = &\pi{\Lambda}
\int d\eta\,d\beta\;\frac{a^3 \sinh\eta}{\tau^3}
\left\{
\frac{(\nabla w)^2}{1-w^2}+(1-w^2)\left((\nabla v)^2+\frac{N^2\tau^2}{a^2
\sinh^2\eta}\right) \right. \nonumber\\
&&+\left.\frac{g_1}{2}\left(\frac{N^2\tau^2}{a^2\sinh^2\eta}(\nabla w)^2+
(\nabla w\times\nabla v)^2\right)\right. \nonumber\\
&&+\left.\frac{g_2}{4}\left[
\frac{(\nabla w)^2}{1-w^2}+(1-w^2)\left((\nabla v)^2+\frac{N^2\tau^2}{a^2
\sinh^2\eta}\right)\right]^2
\right\}\,.
\end{eqnarray}
In toroidal coordinates the gradient includes a factor $a^{-1}$.
Hence the term quadratic in the gradients is proportional to $a$
while the quartic terms are inverse proportional to it.
For soliton solutions, the energy functional has to be varied
with respect to $w, v$ and $a$.

\section{Numerical Results}\label{Nres}

The variational equations for eq.~(\ref{en2}) are highly
nonlinear coupled PDEs and numerically hard to tackle.
Therefore we solved the problem by
a minimization of the energy functional 
which was discretized on an $(\eta,\beta)$ grid. 
The search for the minimum in a high-dimensional space
is feasible using the NETLIB routine $ve08$ with
an algorithm  described in \cite{Gri/Toi:82}.
This method is applicable if the objective function is a sum
$f({\bf x})=\sum f_i({\bf x})$ 
of simpler functions $f_i,$ each of which is
non-constant only for a few components of the (multi-dimensional)
vector {\bf x}.
Thus the Hessian matrix is very sparse and can be updated
locally. This saves a considerable amount of memory and time 
compared to a more naive implementation of a conjugate gradient search.

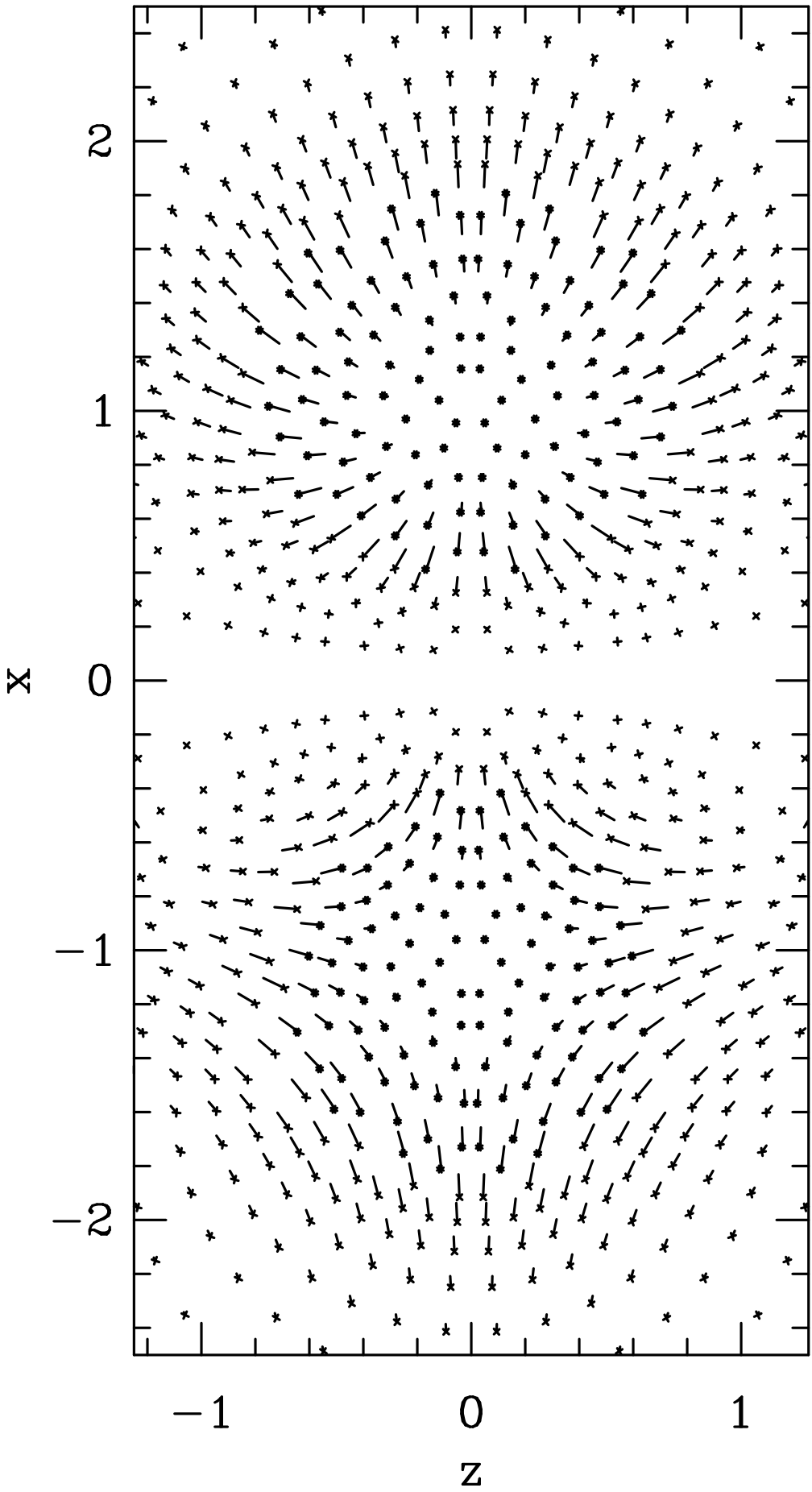
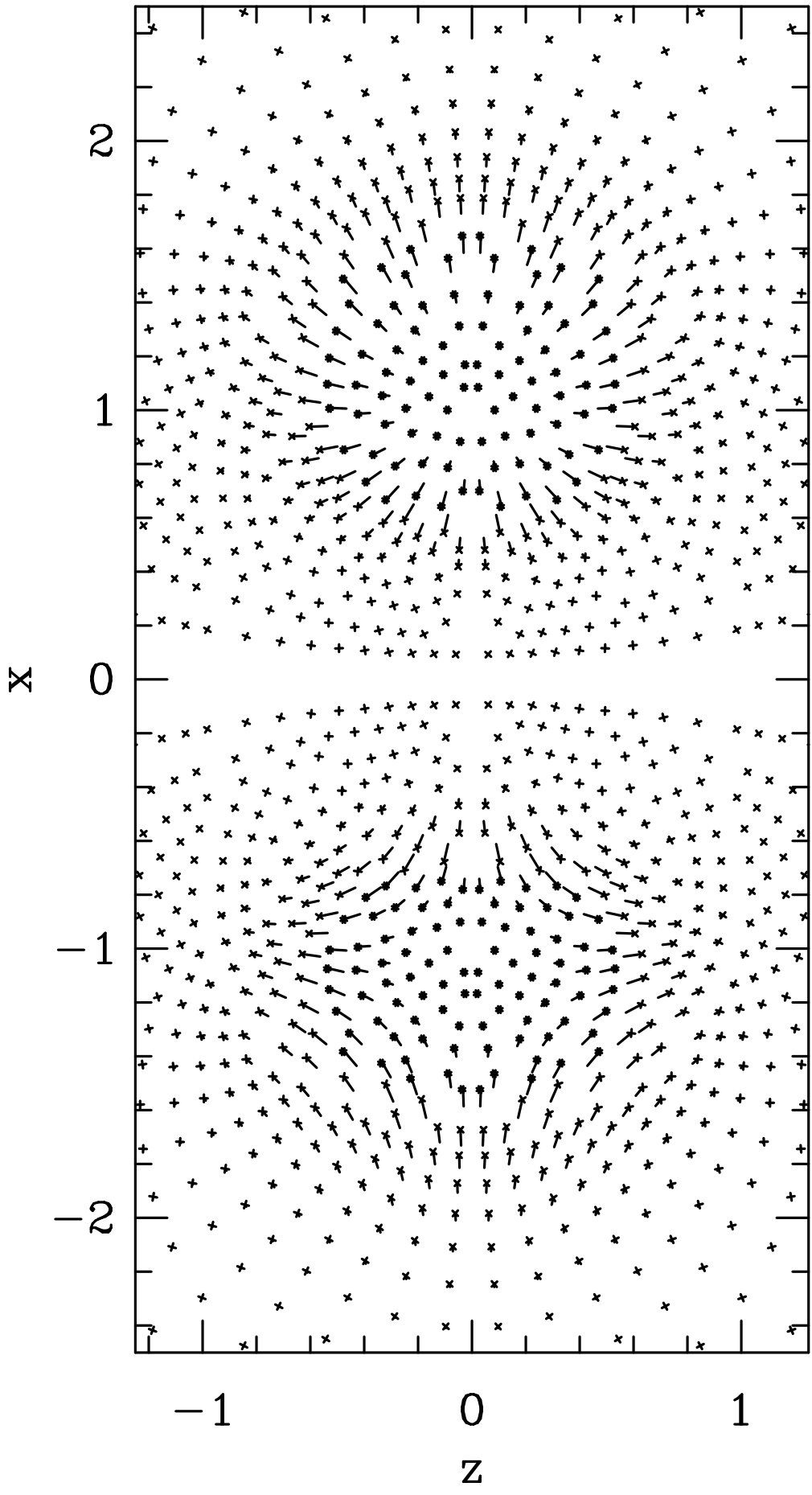
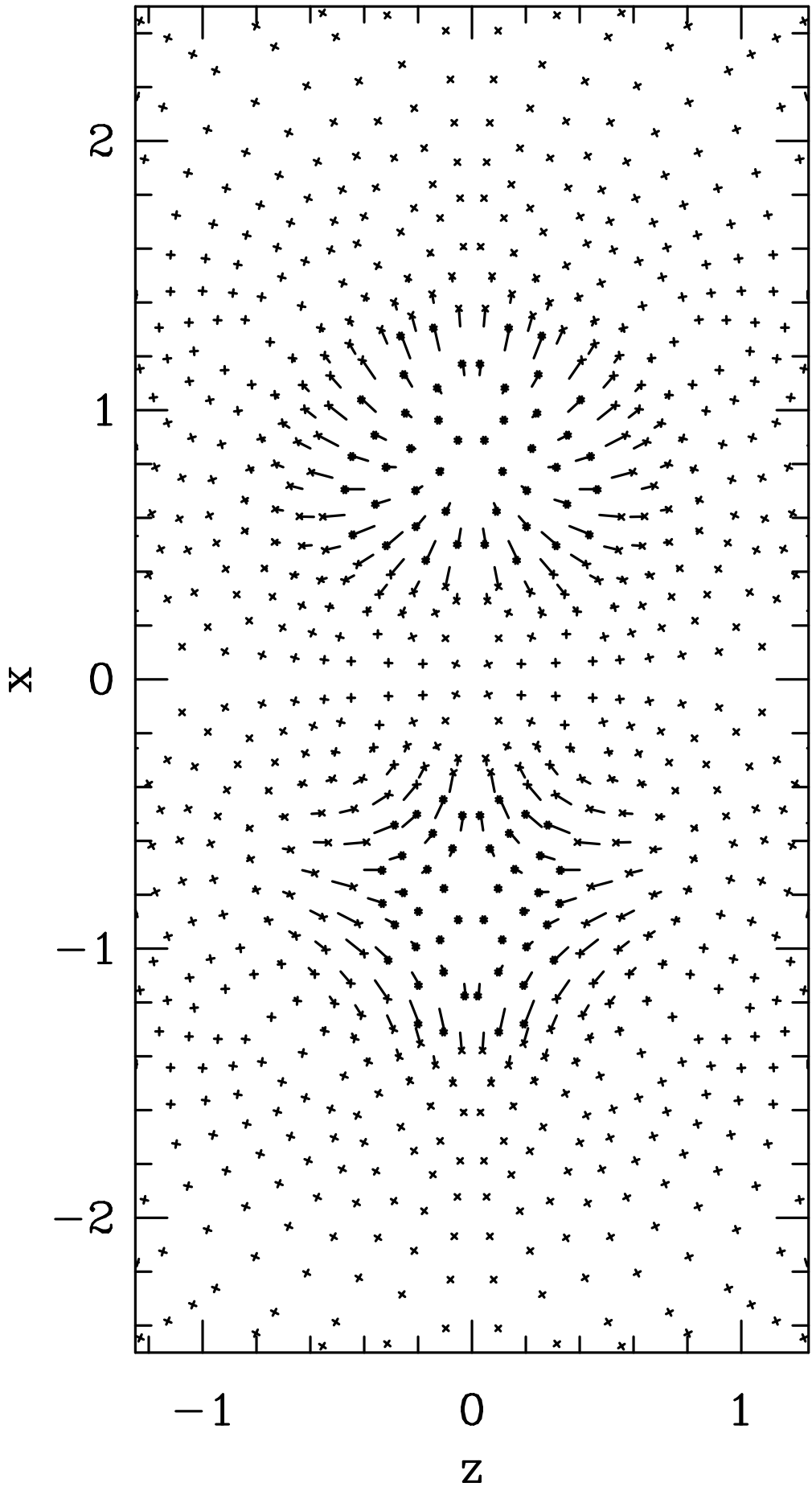
\begin{figure}
\caption{(a) Field configuration in the $xz$-plane for  $H=1,
g_1=0.4, g_2=0.4$.
(b) Field configuration of $H=2, g_1=0.4,g_2=0.4$.
(c) Field configuration with potential term,
$H=1, g_1=1, g_2=0, m=4$. The field is projected into the
$\phi_1\phi_2$-plane. A cross indicates $\phi_3  >0$, a dot
$\phi_3  < 0$. Therefore the vacuum state is denoted by  a cross
only.}
\label{fig1}
\end{figure}

We obtain
field configurations as displayed in
Fig.~\ref{fig1}(a) where the Hopf number equals 1.
In this plot the field $\bbox{\phi}$ is viewed
from above the north pole of target $S^2$.
Iso-vectors in the northern hemisphere terminate in a cross, those
in the southern hemisphere in a dot.
The  toroidal structure of the fields is clearly visible.
Also note that the fields in the southern hemisphere 
span a torus indeed.

There is an interesting interpretation of such configurations
in terms of the 
$O(3)$ $\sigma$-model in (2+1) dimensions, the solutions of which
we call (anti-) baby Skyrmions.
The fields in the positive and negative $x$-halfplane
of Fig.~\ref{fig1}  are  baby Skyrmions and
anti-baby Skyrmions respectively.
This can be understood in the following way.
Wilczek and Zee  \cite{Wil/Zee:83} show that a
(2+1)-dimensional configuration of Hopf number one can be produced
by creating a baby Skyrmion/anti-baby Skyrmion pair from the vacuum,
rotating the (anti-) Skyrmion adiabatically
by $2\pi$ and then annihilating the pair.
In our model time corresponds to the third space dimension,
hence Fig.~\ref{fig1}(a) displays a ``snapshot'' at the time
when the anti-baby Skyrmion is rotated by $\pi$.
Baby Skyrmions are classified by a homotopy invariant
$Q \in {\Bbb Z}$ due to  $\pi_2(S^2) = {\Bbb Z}$.
The  analytic expression for $Q$ is given by

\begin{equation}\label{baby}
Q = \frac{1}{4\pi} \int_{{\Bbb R}^2}\, d{\bf x}\,
\bbox{\phi}\cdot\partial_1 \bbox{\phi}\times \partial_2 \bbox{\phi}\,,
\end{equation}
where 1 and 2 denote cartesian coordinates in ${\Bbb R}^2$.
The topological charge density is half the $\alpha$-component
of {\bf B}. 
The integral over the whole plane vanishes because the contributions
for negative and for positive $x$ exactly cancel.
However, if integrated over the positive
half-plane only, eq.~(\ref{baby}) yields the
baby Skyrmion number for  ansatz ~(\ref{ansatz1}):

\begin{equation}
Q = \frac{1}{8\pi} 
\int_0^{2\pi} d \beta\,\int_0^\infty d\eta\,
\frac{a^2}{\tau^2}B_\alpha = M\,,
\end{equation}
where we use $B_\alpha$ of eq.~(\ref{AB}).

Next we turn to Hop{f}ions of topological charge two.
For  parametrisation eq.~(\ref{ansatz1}) there are two
ways of creating a Hop{f}ion with $H=2$, namely by setting
either $N$ or $M$ to 2.
Both cases correspond to two Hop{f}ions sitting on top of each other.
In order to determine which configuration represents the true
ground state we computed their energies 
and found that the configuration with $N=2, M=1$ yields the
lower energy for all couplings.
The interpretation of the $H=2$ solutions in terms
of a (2+1)-dimensional soliton/anti-soliton pair
is  equivalent to the
one given above for the 1-Hop{f}ion.
Because  the multiplicity of the azimuthal
rotation is $N=2$ for the 2-Hop{f}ion, the
anti-baby Skyrmion
in the negative $x$-halfplane (see Fig.~1(b))
has a relative angle of $\pi$ compared to the anti-baby Skyrmion
of Fig.~1(a).

It is instructive to investigate  how the inclusion of a potential
term $V[\bbox{\phi}]$ alters the configuration.
Its energy can be lowered by rescaling {\bf x} $\to \lambda${\bf x},
($\lambda \to 0$)
under which $V \to \lambda^3  V$.
This means that the potential term induces a ``shrinkage'' of the
configuration in the sense that the favoured position of the
fields is closer to their vacuum value.
This effect 
is counter-balanced by the higher order derivatives
in the energy functional eq.~(\ref{en}).

Any potential explicitly breaks the  models global $O(3)$ symmetry
because $O(3)$ acts transitively on the target space.
We chose
\mbox{$V=m^2\int d{\bf x}\,(1-{\bf n}\cdot\bbox{\phi})$},
where the parameter $m$ is of dimension (length)${}^{-1}$
and, in a quantum version of the theory, becomes
the mass of the elementary excitations.
The minimum energy solution  for $m=4$ can be seen in 
Fig.~1(c).
The tube-like region where the field is in the
southern hemisphere  has clearly shrunk.
Adding a linear potential term  also means that 
the fields fall off exponentially at large distances.
The reason is that the equations of motion become
in the asymptotic limit those of 
the massive Klein-Gordon equation.

The fields of minimal energy correspond, via eq.~(\ref{en}), to 
energy distributions which are displayed in Fig.~2.
Despite the toroidal structure of the fields, we find that
the energy for the Hop{f}ion of $H=1$ is lump-shaped,
see Fig.~2(a).
Although unexpected, this is not entirely unlikely, because the
field changes far more rapidly within the disc 
$\left|{\bf r}\right| \leq a$ than outside it.
Hence the gradient energy can be concentrated in the
vicinity of the origin.

If the potential term becomes very  large compared to the gradient terms
one  expects the energy to become more localized around the
filament where the fields are far away from the vacuum.
We observe this transition to a toroidal energy distribution
at $m \approx 4$ for $g_1=1, g_2=0.$

The energy distribution of the 2-Hop{f}ion  is 
of toroidal shape (for all $m$), as shown in Fig.~2(b).
It is a common feature in many soliton theories that 
solutions of topological charge two are tori, notably 
Skyrmions, baby Skyrmions and  magnetic monopoles.
It is interesting to ask whether the 2-Hop{f}ion is in a 
stable state or likely to decay into two Hop{f}ions of charge one.
As an estimate for short range interactions one can compare the
energy per Hop{f}ion for the solution of $H=1$ and $H=2$
and conclude from the
sign of the energy gap  whether there is a repulsive
or attractive channel.
Our results are plotted in Fig.~3(a), which also
shows the topological bound eq.~(\ref{topbound}).
For a pure Skyrme coupling we obtain 
energies of $197\Lambda$ and $2*158\Lambda$ for the 1-Hop{f}ion and
2-Hop{f}ion respectively.
Moreover, it turns out that for all couplings the 2-Hop{f}ion
has a lower energy per topological unit than the
1-Hop{f}ion.
This indicates that there is a range where the  forces
are attractive and that the 2-Hop{f}ion can be stable at least
under small perturbations. 
Of course, there can be a range in which the forces
are repulsive, however, an investigation of 
such interactions would require a full
(3+1)-dimensional simulation which is beyond our present means.
Also note that the gap between the energies per Hop{f}ion 
is largest when the fourth order terms are purely the Skyrme term.
On the other hand, 
for $g_1 \to -2g_2$, (i.e. $g\to 1)$ the energy of the quartic terms tends to
zero.
Hence the energy of the soliton vanishes 
as a consequence of the above mentioned Hobart-Derrick
theorem.

\section{Spinning Hopfions}\label{Srot}

Finally, we study the effect of a slow rotation
around the axis of symmetry.
For this we use a Lorentz-invariant extension of our
model into (3+1) dimensional space-time.
The energy of the rotating Hop{f}ion $E=E_{rot}+E_{stat}$, where
$E_{stat}$ is the static energy given by eq.~(\ref{en}) and $E_{rot}$ is the
rotation energy functional:
\begin{equation}
E_{rot}\left[ \bbox{\phi} \right] =
\Lambda\int_{{\Bbb R}^3} d{\bf x}\;
\frac{1}{2}
\left(\partial_t \bbox{\phi} \right)^2 +
\frac{g_1}{8}\left(\partial_t \bbox{\phi}  \times
\partial_i \bbox{\phi} \right)^2 +
\frac{g_2}{8}\left(\partial_t \bbox{\phi} \right)^2
\left( \partial_i \bbox{\phi}\right)^2 +
O\left(\left(\partial_t \bbox{\phi}\right)^4\right) \,.
\end{equation}
In the spirit of a moduli space approximation
we assume that the configuration does not alter its shape
due to the rotation (``rigid rotor''), i.e. 
it is given at any time by a static solution  (see \cite{Mak/Ryb:SkMod}
for a review on similar treatment of the Skyrmion).
We impose  time dependence on the
azimuthal angle by \mbox{$\alpha \to \alpha + \frac{\omega}{N} t$} 
with constant velocity $\omega.$
$E_{rot}$ leads to a term  in the energy that is proportional to
$\omega^2$
\begin{equation}
E = E_{stat} + \frac{J}{2} \omega^2\,,
\end{equation}
where  terms $O(\omega^4)$ are neglected.
$J$ is the moment of inertia and, using eq.~(\ref{ansatz1}), given by
\begin{equation}\label{moi}
J = 2\pi\Lambda \int \, d \eta d\beta\,
\left[ 1 + \frac{g_1}{2}\frac{(\nabla w)^2}{1- w^2} + 
\frac{g_2}{2} \left(
\frac{(\nabla w)^2}{1- w^2} + 
\left((\nabla v)^2+\frac{N^2\tau^2}{a^2\sinh^2\eta}\right)(1-w^2)\right)\right]
(1-w^2)\,.
\end{equation}
$J$ can be measured explicitly on the individual solution.
We  plotted the values for $H=1$ and $H=2$ in Fig.~3(b).
The moment of inertia per Hop{f}ion is always larger for the
$H=1$ solution, with an increasing gap for decreasing $g$.
This should be compared with the dependence of
$E_{stat}$ on $g$.

The functional $E_{stat}$ 
(eq.~(\ref{en}))  is invariant under $\alpha$-rotations
while the fields of ansatz~(\ref{ansatz1}) are clearly not.
Therefore, upon quantization, 
the coordinate $\alpha$ describes a zero-mode and
requires treatment as a collective coordinate.
This is similar to the problem of the rotating radially symmetric 
Skyrmion.
In analogy to the Skyrme model we 
therefore use, as a first approximation, the spectrum obtained
by a straightforward quantization.
The canonical momentum is $l = i\frac{d}{d\alpha}, (\hbar=1)$ and 
the rotational energy $E_{rot} =- l^2/2J$.
It is then trivial to solve the eigenvalue problem 
$E_{rot}\psi = \lambda \psi$, which gives 
$\lambda_n = \frac{n^2}{2J}$.

\section{Conclusions}\label{Conc}

We have studied topological solitons in a generalized non-linear 
$O(3)$ $\sigma$-model in three space dimensions. 
Physically one may think of them as a model 
for hadronic matter or
topological defects in a condensed matter system.
By using a general 
ansatz for the fields we
obtained explicit numerical solutions for soliton number one and two.
Unexpectedly, the energy of the 1-Hop{f}ion is distributed as a lump.
We also observed that two solitons sitting on top of each other 
attract, thus indicating a stable configuration.

There are several interesting questions which remain unanswered.
In particular, the stability of Hop{f}ions of higher topological charge
deserves some scrutiny. It is worthwhile asking 
how multi-solitons which sit on top of each other,
or at least are very close, behave under induced perturbations.
In analogy to planar $O(3)$ $\sigma$-models 
there might be several decay channels 
into less symmetric configurations \cite{Piet/Schr:95}.

At the opposite end of the scale, it would be instructive to look in
greater detail at the interaction potential of two or more
well-separated Hop{f}ions. 
This is also interesting in comparison to the well-studied 
dynamics of Skyrmions and monopoles. 
Clearly, a first step in such an investigation would be to determine
the asymptotic fields of the Hopf soliton. It seems obvious that 
inter-soliton forces will depend on the orientation of the
Hop{f}ions.

The complete description of  Hop{f}ion dynamics
would require a huge numerical effort which can, however,  
possibly be reduced by an appropriate approximation scheme.
For Bogomol'nyi solitons, the low-energy behaviour  can be approximated
via the truncation of the dynamics to the moduli space.
Although our numerical results show that
Hop{f}ions are not of Bogomol'nyi type, given that the static forces
between them are weak, there is a chance that  
their dynamics can be described by 
some kind of moduli space approximation, in analogy to 
Skyrmions (which are also not of Bogomol'nyi type).

Finally, it seems  worth to study spinning 
Hop{f}ions in a more sophisticated way.
This should include an assessment of the back reaction of the 
rotation on the matter fields. 
From this one expects a non-trivial shift of the energy levels in the
rotation spectrum and 
possibly radiation of excessive energy.

\acknowledgements
It is a pleasure to thank Wojtek Zakrzewski, Jacek Dziarmaga and
Rob deJeu for helpful discussions.
We also wish to
thank Bernd Schroers for making reference \cite{WaK} available to
us.
JG acknowledges an EPSRC grant No.~94002269. 
MH is supported by  
Deutscher Akademischer Austauschdienst.


\newpage

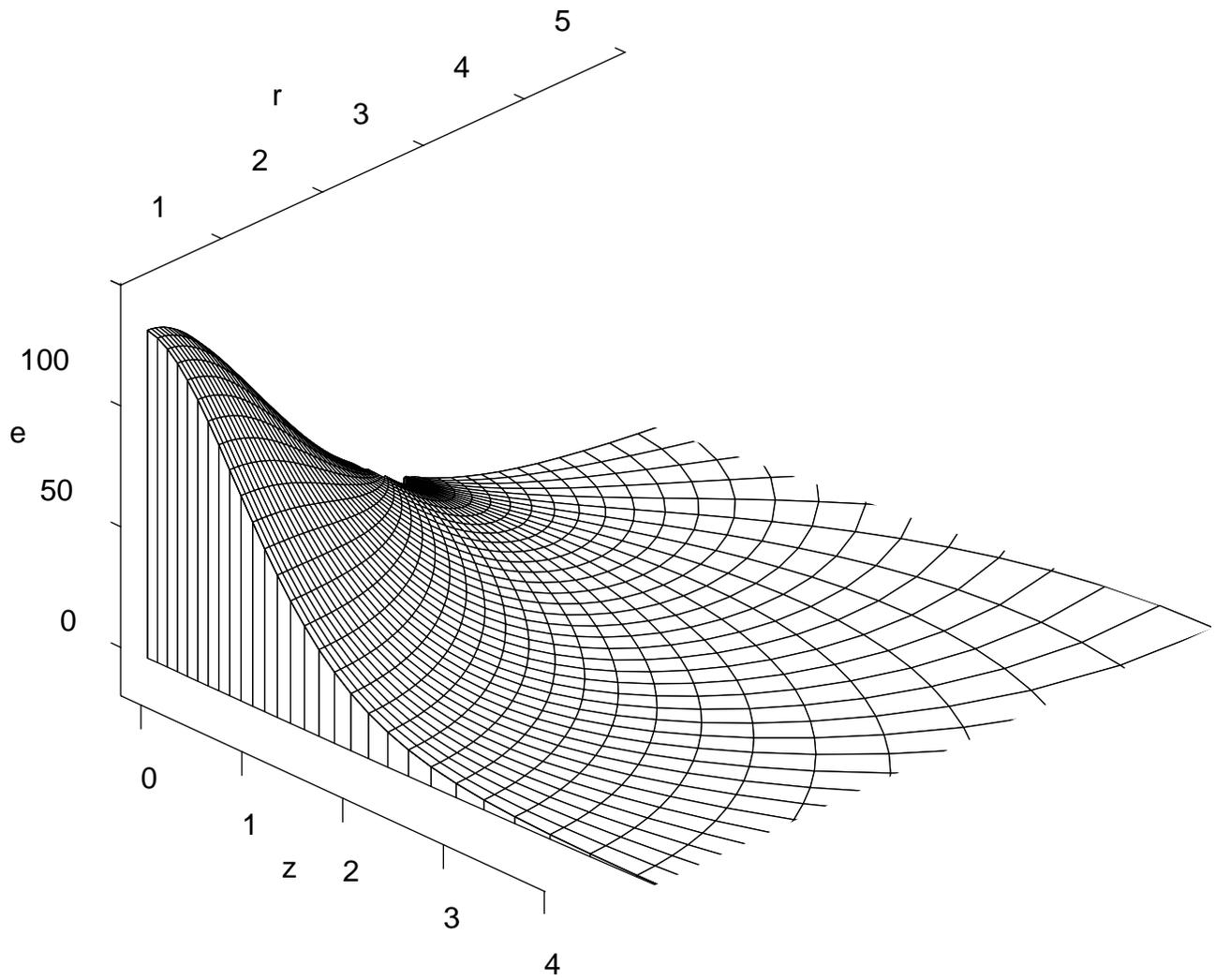
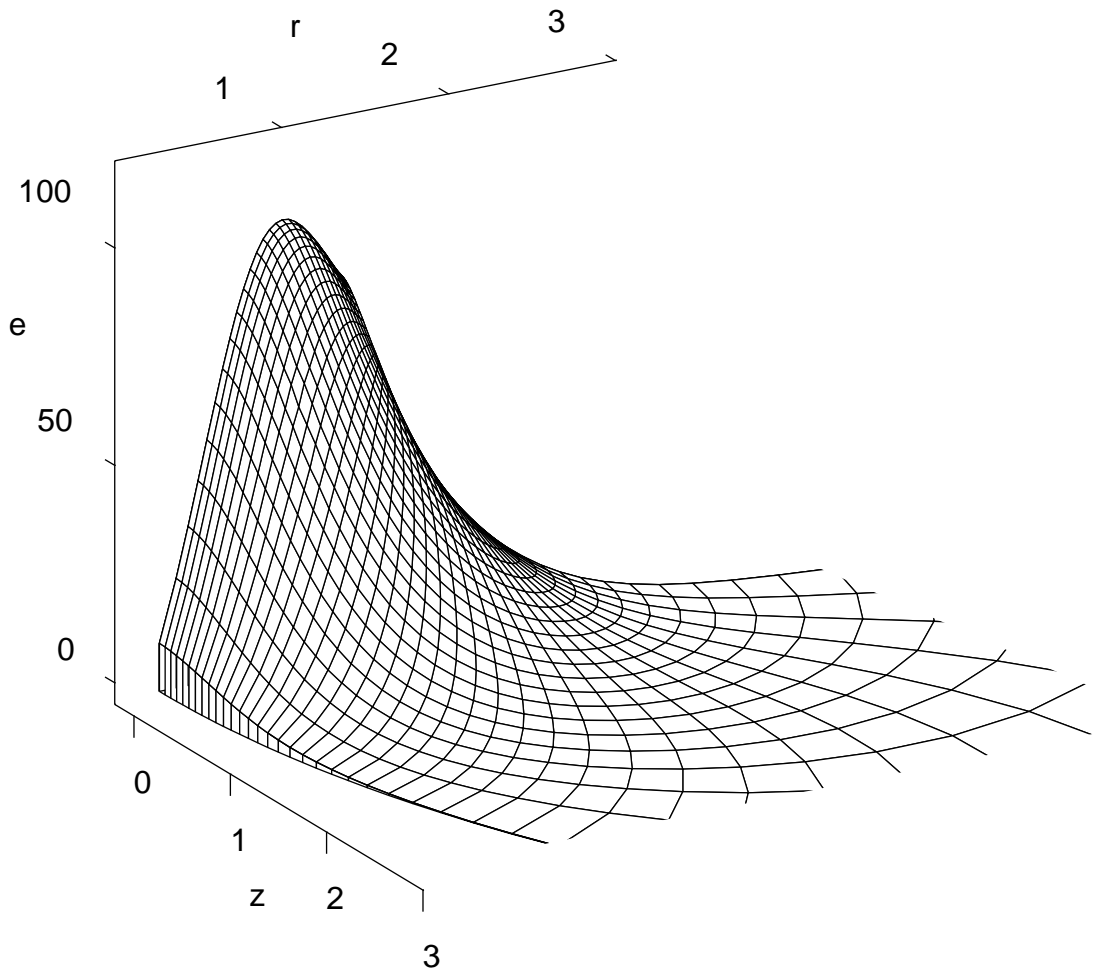
\begin{figure}
\caption{(a) Energy density $e$ (arbitrary units) 
 for $H=1, g_1=0.4, g_2=0$ 
in cylindrical coordinates $r,z$.
(b) Energy density $e$ for $H=2, g_1=0.4, g_2=0.8$ over  $r,z$.}
\label{fig2}
\end{figure}

\def\bbox{\relax} 

\begin{figure}
\caption{(a) Dependence of the energy $E_{stat}$ per Hopfion
on the quartic couplings. They are parametrized as $g_1=1-3g,
g_2=g.$ Hence $g=0$ corresponds to pure 
$(\partial_i \bbox{\phi}  \times
\partial_j \bbox{\phi})^2$ coupling, $g=1/3$ to pure 
$(\partial_i \bbox{\phi} )^2( \partial_j \bbox{\phi})^2$
coupling and $g=1$ to the case
$(\partial_i \bbox{\phi} )^2( \partial_j \bbox{\phi})^2 -2
(\partial_i \bbox{\phi}  \times
\partial_j \bbox{\phi})^2.$ 
The energy is given in units of $\Lambda.$
The topological bounds for pure Skyrme coupling are also displayed.
(b) Dependence of the 
moment of inertia $J$ 
per Hop{f}ion on the coupling $g$.}
\label{fig3}
\end{figure}
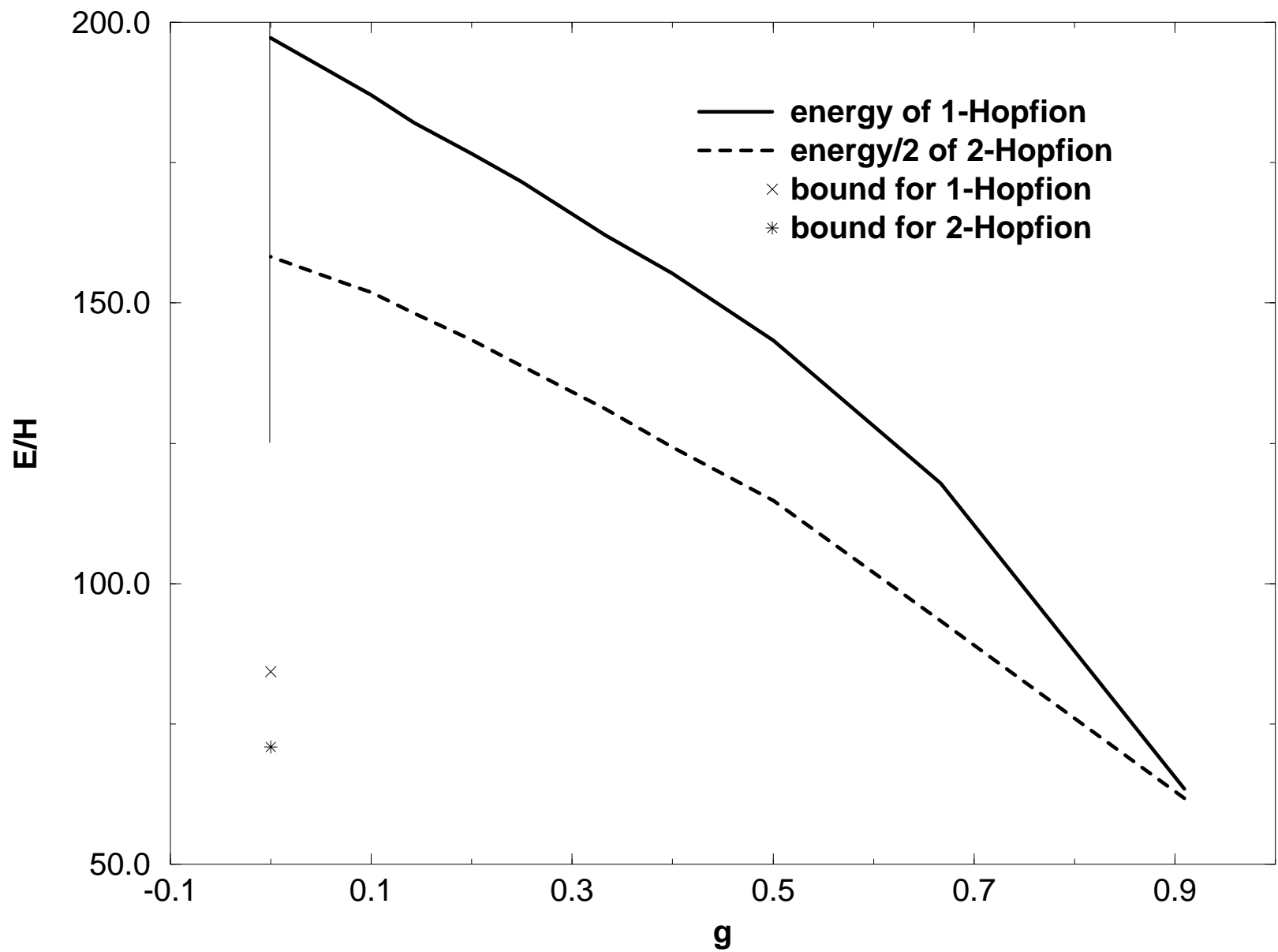
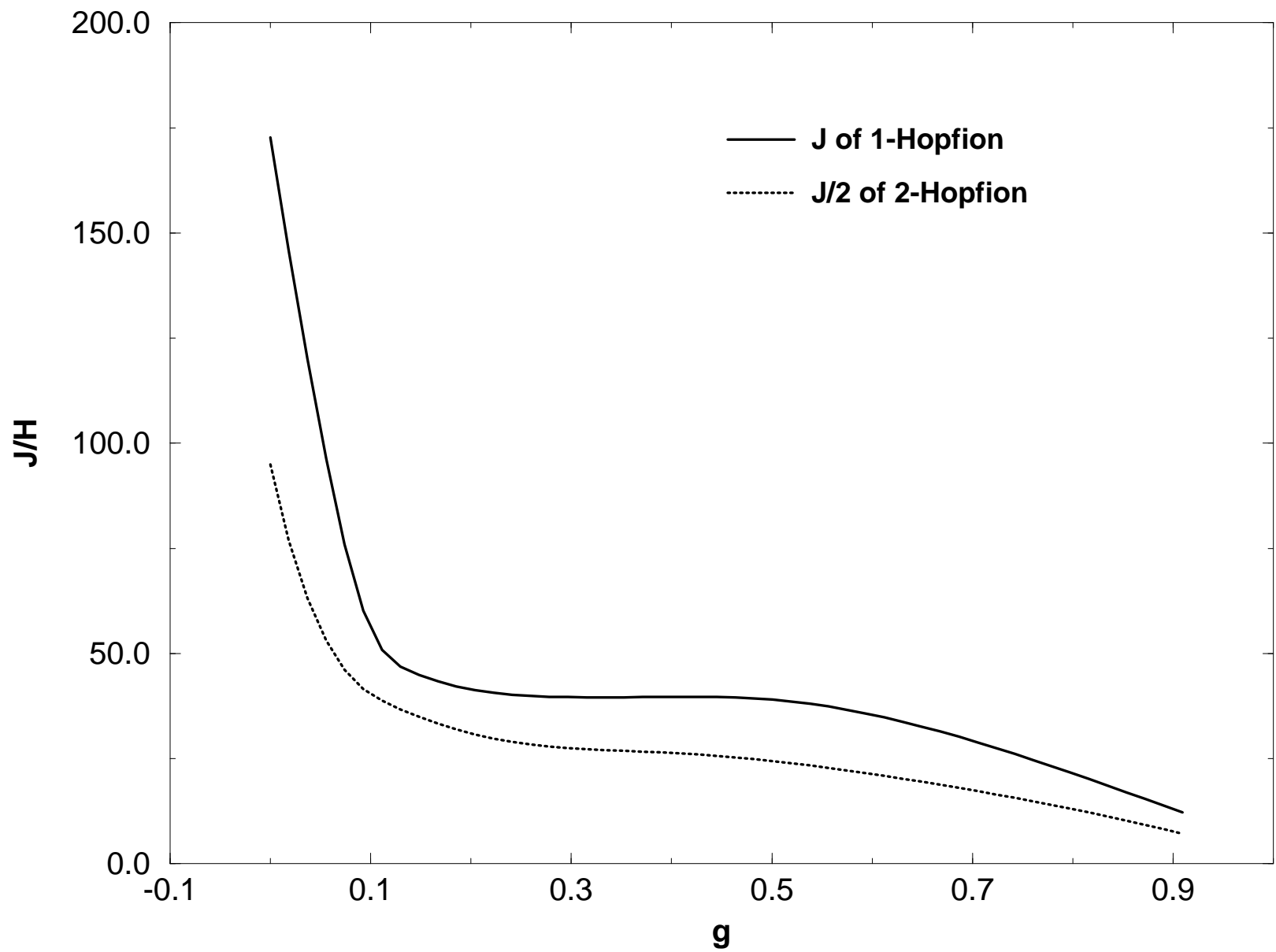
\end{document}